\documentclass{jkas}

\def\year{2026} 
\def\volume{TBD} 
\def\issue{TBD} 
\def\beginpage{1} 
\def\received{---} 
\def\accepted{---} 
\def\published{---} 
\date{Received \received; Accepted \accepted; Published \published}

\newcommand\fdg{\mbox{$.\!\!^\circ$}}%
\title{%
Kinematics and Dynamics of the Open Cluster NGC 2302
}

\author[1,$\dagger$]{Yuna Lee}{}
\author[2]{Jongsuk Hong}{}
\author[1,$\star$]{Beomdu Lim}{}

\affil[1]{Department of Earth Sciences Education, Kongju National University, 56 Gongjudaehak-ro, Gongju-si, Chungcheongnam-do 32588, Republic of Korea}
\affil[2]{Korea Astronomy and Space Science Institute, 776 Daedeok-daero, Yuseong-gu, Daejeon 34055, Republic of Korea}

\def\corrauthor{%
B. Lim, \email{blim@kongju.ac.kr}
}

\def\runningauthor{%
Lee et al.
}

\def\runningtitle{%
Kinematics and Dynamics: NGC 2302
}

\def\keywords{%
Open clusters : individual NGC 2302 --- Stellar kinematics --- Stellar dynamics
}

\def\abstracttext{
Open clusters are ideal observational testbeds to understand 
the dynamics of stellar systems. We present a dynamical study of the young 
open cluster NGC 2302. The latest Gaia data and $UBVIJHK_s$ photometric data 
are used in this study. A total of 117 stars are selected as the genuine 
members using the Gaia data. This cluster is, on average, reddened 
by $\langle E(B-V) \rangle = 0.24 \pm 0.06 (s.d.)$. The ratio of total-to-selective extinction 
($R_V$) in the direction of NGC 2302 is $2.8 \pm 0.1$. The cluster 
distance is determined to be $1.16 \pm 0.08$ kpc using Gaia parallaxes. Theoretical 
isochrone fitting for $Z = 0.008$ on color-magnitude diagrams yields 
an age of $80 \pm 20$ Myr. The relative proper motions of individual members 
show no significant radial expansion or contraction. NGC 2302 contains a total 
stellar mass of $333 \pm 48 \ M_{\odot}$. The one-dimensional velocity dispersion 
is approximately 0.26 km s$^{-1}$, which is comparable to the viral velocity dispersion 
of 0.27 km s$^{-1}$ derived from its total mass. 
Its relaxation time is estimated to be approximately 
90 Myr, which is similar to the age of the cluster within 
the uncertainty in age estimation. Finally, we report a pattern 
of mass segregation in the radial distribution of stellar 
masses. Our results suggest that NGC 2302 is virialized and currently approaching a state of dynamical relaxation. However, because no definitive evidence of kinetic energy equipartition is found, the possibility of the in-situ formation of high-mass stars within the central region should be carefully considered.
}

\begin{document}
\jkashead 

\section{Introduction}
Open clusters are stellar systems composed of several hundreds 
to thousands of coeval stars in a wide range of masses. They 
are, thus, ideal laboratories to examine the theory of stellar 
evolution. Most stars form in open clusters or stellar associations 
in the Galactic disk \citep{lada2003,porras2003}, so that such young 
stellar systems are tracers of star-forming regions distributed 
along spiral arms in the Galactic plane \citep{Cantat-Gaudin2018}. 

Open clusters are also useful objects to test the 
theory of stellar dynamics \citep{binney2008}. Recent 
Gaia data \citep{gaia2016,gaia2023} have promoted the 
study of their dynamical evolution. In addition, these 
clusters contribute to field stellar population because 
most of them are expected to be dissolved into the 
Galactic disk \citep{lada2003}. Gas expulsion, the 
encounter with giant molecular clouds, and Galactic 
tidal field are responsible for the dissolution of 
open clusters \citep{gieles2006,moeckel2010}.

According to the cluster mass function, low-mass stellar 
clusters are born in far greater abundance than their massive 
counterparts \citep{portegies2010,just2023}. However, due to their 
shallower potential wells, only a small fraction of these low-mass 
systems are expected to survive early gas expulsion and tidal disruption 
to remain as gravitationally bound clusters \citep{lada2003}. 
Understanding the detailed dynamical evolution of these low-mass 
clusters is therefore crucial for constraining cluster disruption 
mechanisms and the overall origin of field star populations. 
Nevertheless, they have received considerably less attention 
compared to more massive clusters.

In this study, we investigate the low-mass 
open cluster NGC 2302, located in the Galactic 
anticenter. This cluster has been included 
as a target of extensive surveys based on 
the Gaia data \citep{costa2015,bossini2019,cantat-gaudin2020,
almeida2023,hunt2023}, but its fundamental 
parameters and the state of dynamical evolution 
are poorly studied. According to the previous 
studies, NGC 2302 is located at a distance 
of approximately 1.16 -- 1.40 kpc from the Sun. 
The age of the cluster was estimated to be 
about 100 Myr.

The goals of this study are to determine the fundamental parameters of NGC 2302 using a homogeneous set of $UBVIJHK_SG_{\mathrm{BP}}G_{\mathrm{RP}}$ photometry and to investigate its dynamical state. The remainder of this paper is structured as follows. Section~\ref{sec2} describes our imaging observations and the archival data used in this work. Section~\ref{sec3} details the membership selection criteria. In Section~\ref{sec4}, we present the derived fundamental parameters and analyze the cluster dynamics. Finally, a summary of our primary results is provided in Section~\ref{sec6}.

\section{Data} \label{sec2}
\subsection{Imaging observation}
\begin{figure}
    \centering
    \includegraphics[width=80mm]{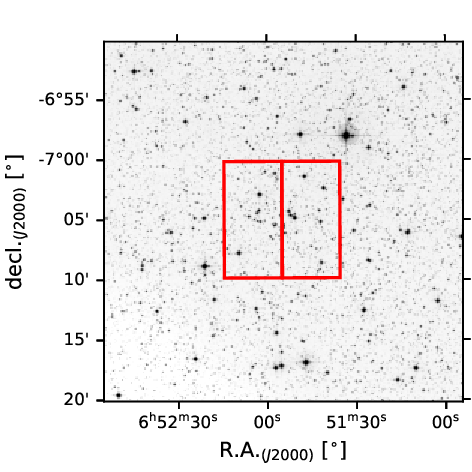}      
    \caption{Digital Sky Survey 2 image for a 30$^\prime \times$30$^\prime$ box region centered on NGC 2302 ($\alpha = 6^\text{h}51^\text{m}54_\cdot^\text{s}5$, $\delta = -7^\circ05^\prime09_\cdot^{\prime\prime}6$). Red lines represent the observed field, conducted with the Mont4k CCD camera attached onto the Kuiper 61$^{\prime\prime}$ telescope; note that the field appears as two adjacent rectangles due to the detector layout consisting of two $2\text{K} \times 4\text{K}$ CCD chips with a small inter-chip gap ($\sim 5\text{ pixels}$ or $\sim 2^{\prime\prime}$).}
    \label{fig1}
\end{figure}

We performed imaging observation of NGC 2302 
on 2011 November 2, using $UBVI$ filters and 
the Mont4k CCD camera attached to the Kuiper 
61$^{\prime\prime}$ telescope ($f$/13.5) on Mt. 
Bigelow in Arizona, USA. For photometric 
calibration, a number of standard stars 
\citep{menzies1991,landolt1992,KvR98} were 
also observed on the same night. The Mont4k CCD camera 
covers a 9$_.^\prime$7$\times$9$_.^\prime$7 (Figure~\ref{fig1}). 
All images were taken in a $3\times3$ binning mode. 
Our observation is summarized in Table~\ref{tab1}. 

The images were reduced using the IRAF/CCDRED 
package. This pre-processing includes bias 
subtraction, flat fielding, and correction for 
shutter shading (see \citealt{sos4}). We 
detected sources in the observed images by means 
of the IRAF/DAOFIND task. Point spread function 
photometry was conducted for the detected sources 
using the IRAF/DAOPHOT package. Several spurious 
sources were excluded through visual inspection 
of individual images. We obtained photometric 
data of 1491 sources in total.

\begin{figure}[h]
    \centering
    \includegraphics[width=80mm]{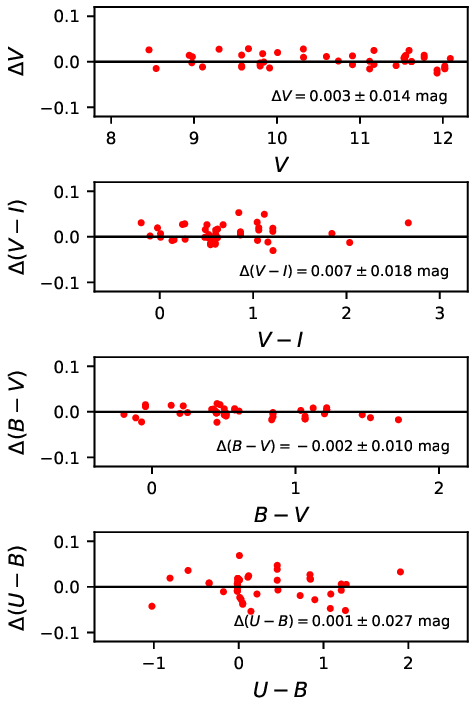}      
    \caption{ Photometric comparison between the standard stars and our observed stars. Red dots represent the residuals for individual stars, with the mean value and standard deviation ($\sigma$) labeled at the bottom of each panel. }
    \label{fig2}
\end{figure}

The instrumental magnitudes of individual sources 
were transformed to the standard magnitudes tied 
to the Johnson-Cousins system \citep{johnson1953,B90} 
using the transformation relations of \citep{sos4}. 
We present the atmospheric extinction coefficients 
and photometric zero points of all passbands in 
Table~\ref{tab2}. A photometric study of NGC 2302 
was conducted as part of a large survey by \citet{costa2015}; 
however, the photometric data for this cluster were 
not published, which makes it impossible to check for 
photometric consistency. We thus transformed the 
instrumental magnitudes and colors of the observed 
standard stars to the standard ones using the same 
transformation relations and coefficients as above. 
Figure~\ref{fig2} compares the two data sets. The 
difference between the two data sets is less than 1\% 
in the zero points, with a scatter of 2–3\%. Our 
photometric data are well tied to the Johnson-Cousins 
system.

\begin{table}[t]
\caption{Observation log\label{tab1}}
\centering
\begin{tabular}{lccc}
\toprule
Target & Date & Filter & Exposure Time [s] \\
\midrule
NGC 2302 & 2011 Nov. 2 & $I$ & 5 and 120 \\
          &    & $V$ & 5 and 180 \\
        && $B$ & 7 and 300\\
        && $U$ & 15 and 600 \\
\bottomrule
\end{tabular}
\end{table}

\begin{table}[t]
\caption{Photometric coefficients\label{tab2}}
\centering
\scriptsize
\setlength{\tabcolsep}{4pt}
\begin{tabular}{lcccc}
\toprule
Filter & $k_1$ & $k_2$ & $\alpha$ & $\zeta$   \\
\midrule
$U$ & $0.438\pm0.006$ & $0.026\pm0.004$ & 0.000 & $22.073\pm0.012$ \\
$B$ & $0.238\pm0.004$ & $0.022\pm0.002$ & $0.009\pm0.001$ & $23.554\pm 0.009$ \\
$V$ & $0.133\pm0.004$ & 0.000 & $0.007\pm 0.001$ & $23.573\pm0.009$ \\
$I$ & $0.053\pm0.004$ & 0.000 & 0.000 & $22.214\pm0.011$ \\
\bottomrule
\end{tabular}
\end{table}

\subsection{Catalogs}
Gaia Data Release 3 (Gaia DR3; \cite{gaia2023}) 
provides highly accurate astrometric and photometric 
data. For this study, we collected Gaia parallaxes, proper 
motions (PMs), radial velocities, and photometric 
data for stars within a $1^{\circ} \times 1^{\circ}$ field 
centered at (R.A.$_{2000}$, decl.$_{2000}$) $\approx$ 
($102\fdg98934$, $-7\fdg08672$). To ensure the significance 
of our results, we restricted our analysis to stars brighter 
than $G_{RP} < 18$ with parallaxes greater than 
three times the associated error ($\pi > 3 \cdot\varepsilon(\pi)$). 
Finally, the zero-point offsets for the parallaxes 
were corrected using the public Python code provided by 
\citet{lindegren2021}. This catalog contains 60,273 stars 
in total and was used as a master catalog in this study.

We also downloaded the near-infrared photometric data for 
stars within the same field of view as our observations from 
the Two Micron All Sky Survey (2MASS; \citealt{skrutskie2006}). 
Stars with high-quality of photometry flagged by `AAA' were 
used to transform the pixel coordinates of our optical 
photometric data into the equatorial coordinates at epoch J2000. 
Combined with our optical data, the near-infrared photometry is 
useful for investigating the reddening law in the direction 
of the cluster.

Our photometric data and the 2MASS data were 
matched the stars in the master catalog within 
a radius of $1^{\prime\prime}$. Consequently, a total of 
1386 stars from our data and 236 stars from 
the 2MASS data were found to have counterparts 
in the master catalog.


\begin{figure*}
    \centering
    \includegraphics[width=160mm]{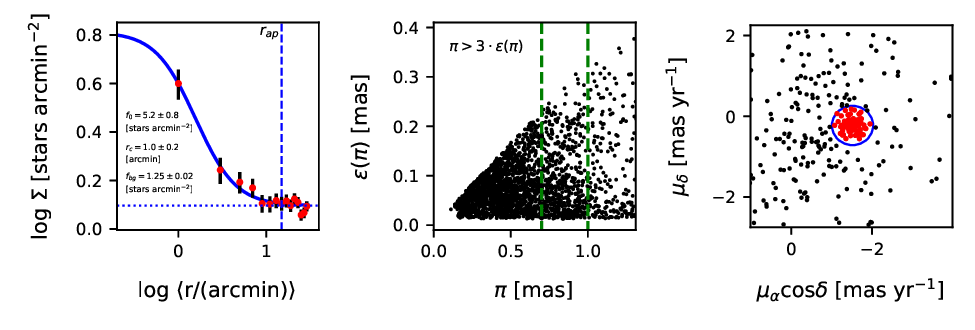}     
    \caption{Structural and kinematic properties of NGC 2302 for membership selection.
Left panel : Radial surface density profile of the NGC 2302 field. The observed stellar densities and their corresponding Poisson errors are denoted by red dots and black vertical bars, respectively. The solid blue curve represents the best-fit King \citep{king1962} profile, and the horizontal dotted line indicates the mean background surface density. The vertical dashed line marks the adopted apparent radius ($r_{\mathrm{ap}} = 15^{\prime}$).
Middle panel : Parallax vs. parallax error distribution for stars located within $r_{\mathrm{ap}}$ from the cluster center and satisfying $\pi > 3\epsilon(\pi)$. The two dashed green lines delineate the adopted parallax range for cluster members.
Right panel : PM distribution of stars within the selected parallax range. Red dots represent high-probability member candidates located within the $3.5\sigma$ ellipse (blue), where the center and axes correspond to the mean PM and 3.5 times the standard deviations in right ascension and declination, respectively. Excluded stars are indicated by black dots. }
    \label{fig3}
\end{figure*}

\begin{figure}[tb]
    \centering
    \includegraphics[width=80mm]{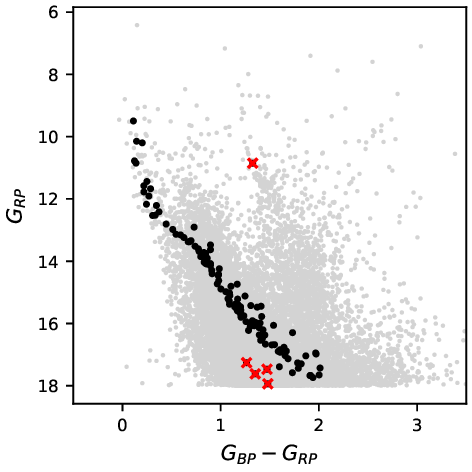}
    \caption{CMD of NGC 2302. Black dots represent cluster member candidates selected based on the Gaia astrometric data. Grey dots denote field interlopers excluded during the astrometric membership selection. Crosses indicate stars that were further eliminated because their positions on the CMD cannot be accounted for by the stellar evolution of coeval stars. }
    \label{fig4}
\end{figure}
\section{Member Selection} \label{sec3}
Since open clusters are a representative 
of disk stellar population, the observed 
images include a large number of disk stars 
as well as cluster members. For this reason, 
member selection is the most important 
procedure to derive the physical parameters 
of open clusters. Members of a given cluster 
have coeval stars sharing similar kinematics 
and chemical composition at the almost the 
same distance. These points allow us to 
reliably select genuine members. 

We first derived the radial surface density profile 
of stars to confine a 
cluster region. This procedure helps 
us to exclude a large number of field 
interlopers. To do this, we counted 
the number of stars within concentric 
rings with a width of 2$^{\prime}$. The 
number of stars in a concentric ring was 
then divided by the associated area, 
yielding the surface density of a given 
ring. The Poisson noise normalized by 
the area of each ring was adopted as 
the uncertainty in surface density. 

The left panel of Figure~\ref{fig3} 
displays the the surface density profile 
of NGC 2302. The surface density smoothly 
drops and is comparable to the number 
density of field stars at around 
$r \sim 15^{\prime}$. We adopted the central 
distance as the apparent radius ($r_{ap}$) 
of this cluster. Hereafter, we searched 
for members within this radius.

The middle panel of Figure~\ref{fig3} 
exhibits the parallax distribution of 
stars. Since members are at the almost 
the same distance, they are likely to be found 
in a narrow range of parallax. In the 
plot, it was challenging to identify 
the overdense region associated with the 
cluster because of its intrinsically 
low surface density compared to that 
of distant stars. We attempted to identify 
the cluster members adjusting the 
limiting magnitudes of the sample and 
confined the parallax range from 0.7 mas 
to 1.0 mas. This range includes the distances 
(1.3 --1.4 kpc) determined by previous studies 
\citep{costa2015,cantatanders2020}.

Stars within the parallax range are plotted 
in the right panel of Figure~\ref{fig3}. 
A group of stars is concentrated within narrow 
ranges of PMs ($\mu_{\alpha}\cos\delta$, $\mu_{\delta}$), 
suggesting they are likely member candidates. 
We selected the most probable members using an 
iterative sigma-clipping method. Initially, we 
computed the mean and standard deviation of the 
PMs. Stars with PMs deviating more than 3.5$\sigma$ 
from the mean were excluded. The mean and standard 
deviation were then recomputed, and this procedure 
was repeated until the values converged. Since the cluster 
members are heavily concentrated within a region much 
narrower than this $3.5\sigma$ threshold, this criterion 
covers a sufficiently wide range of proper motions to act 
purely as a statistical decontamination, rather than 
imposing any dynamical constraints.

Figure~\ref{fig4} presents the color-magnitude 
diagram (CMD) of the member candidates. Most 
candidates align along a well-defined main 
sequence, as expected from a coeval stellar 
population. Nevertheless, some field interlopers remain. 
One red giant star was initially selected as a member, 
but its luminosity is lower than that of the brightest 
main sequence main sequence stars -- a discrepancy 
that cannot be explained by coeval stellar evolution. 
Furthermore, several of the faintest stars exhibit 
bluer colors than other candidates at comparable magnitudes. 
Given that their color errors are smaller than the typical 
uncertainties for stars fainter than 17 in $G_{RP}$, this 
color spread is unlikely to originate from measurement 
errors. Consequently, we excluded both the red giant and 
these faint blue stars from our final sample. In total, 117 
stars were identified as genuine members of NGC 2302.

\begin{figure}
    \centering
    \includegraphics[width=80mm]{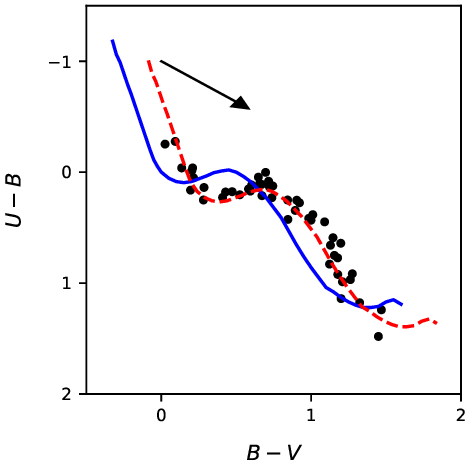}      
    \caption{Color-color diagram of the NGC 2302. The solid and dashed lines represent the intrinsic color-color relation \citep{sos0} and the same relation reddened by $E(B-V) = 0.24$, which corresponds to the mean reddening value derived from six early-type members. The arrow indicates the reddening vector on this diagram. }
    \label{fig5}
\end{figure}

\begin{figure*}
    \centering
    \includegraphics[width=165mm]{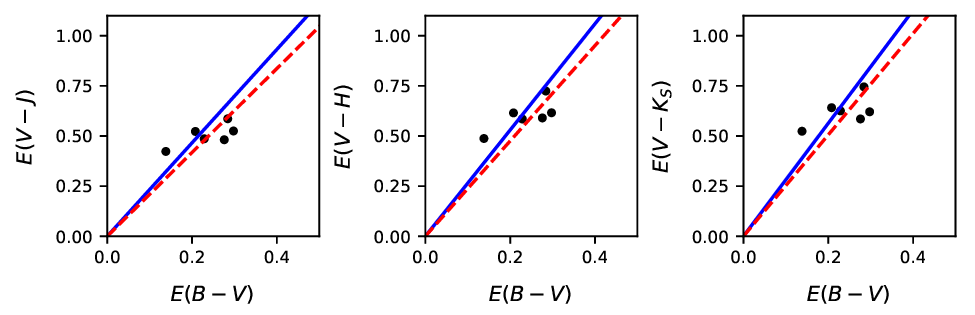}      
    \caption{ Color excess ratios, $E(V-\lambda)$ versus $E(B-V)$, for the six early-type members (black dots), where $\lambda$ represents the 2MASS $J$, $H$, and $K_S$ bands. The red solid line indicates the best-fit linear relation yielding $R_V = 2.8$, whereas the blue line denotes the standard reddening law with $R_V = 3.1$.} 
    \label{fig6}
\end{figure*}


\section{Fundamental Parameters} \label{sec4}

\subsection{Structure}\label{sec4.1}
The apparent radius of NGC 2302 was determined to be $15^{\prime}$, which is equivalent to 5.1~pc at a distance of 1.16~kpc (see Section~\ref{sec4.3}). The number density of stars within this radius is $1.5 \pm 0.1 \ \mathrm{stars} \ \mathrm{pc}^{-2}$. We also determined its half-number radius to be approximately 1.9~pc. Within this half-number radius, the mean number density increases to $5.1 \pm 0.7 \ \mathrm{stars} \ \mathrm{pc}^{-2}$, which is 3.4 times higher than the total average density.

To better characterize the structural properties and quantify the central concentration of the cluster, we fit the observed surface density profile to the empirical King density model \citep{king1962}:\begin{equation}\Sigma (r) = f_{\mathrm{bg}} + \frac{f_0}{1+(r/r_c)^2},\end{equation}where $f_{\mathrm{bg}}$, $f_0$, and $r_c$ represent the background stellar density, central surface density, and core radius, respectively. The best-fit model and its corresponding parameters are presented and labeled in the left panel of Figure~\ref{fig3}. The derived core radius of this cluster is $r_c = 0.35 \pm 0.05$~pc. This exceptionally small core radius relative to both the half-number radius and the overall cluster radius firmly supports the presence of a strong central concentration in NGC 2302, compared to typical open clusters \citep{KPS13}.
  
\subsection{Reddening and Reddening law} \label{sec4.2}
The reddening of stars are, in principle, determined 
by comparing their observed colors with the intrinsic 
values. As intrinsic color relations and reddening 
vectors are well-established for early-type stars \citep{sos0}, 
we determined the individual reddening of six early-type members 
($B - V < 0.3$ and $U - B < 0.1$) by adopting the intrinsic 
color relation and reddening vector on the $(U - B, B - V)$ diagram 
(Figure~\ref{fig5}). Following \citet{sos0}, 
we adopted a reddening vector of $E(U-B)/E(B-V) = 0.72 + 0.025E(B-V)$. 
The reddening values for these members range from 
0.14 to 0.30, with a mean of $\langle E(B-V) \rangle =  0.24\pm 0.05$ 
(s.d.). This result is in a good agreement with 
previous studies \citep{kharchenko2009,costa2015}. 
Furthermore, the standard deviation is comparable 
to the photometric errors, suggesting that 
differential reddening across the cluster is negligible.

To compute the total extinction in the $V$ band, 
adopting an appropriate ratio of total-to-selective 
extinction, $R_V$, is essential. We examined 
the reddening law toward this cluster by combining 
optical and near-infrared photometry of the six 
early-type members. The intrinsic colors of these stars, 
$(V - \lambda)_0$, were obtained by interpolating 
their $(B-V)_0$ values to the intrinsic color 
relations between $(B-V)_0$ and $(V - \lambda)_0$ 
provided by \citet{sos0}, where $\lambda$ represents 
the $J$, $H$, and $K_S$ bands. The color excess 
was then calculated by subtracting the intrinsic color 
from the observed one.

Figure~\ref{fig6} exhibits the color excess 
ratios, $E(V - \lambda)/E(B-V)$. We determined 
the slope between color excesses using the 
least-squares method. $R_V$ values were then 
calculated using the Equations (5) -- (8) from 
\citet{sos0}. The resulting mean $R_V$ is $2.8 \pm 0.1$; 
notably, all three different domains consistently 
indicate that the $R_V$ toward this cluster 
is smaller than the Galactic mean of 3.1 \citep{FM07},

\subsection{Distance, Age, and Total Mass} \label{sec4.3}
\begin{figure}
    \centering
    \includegraphics[width=80mm]{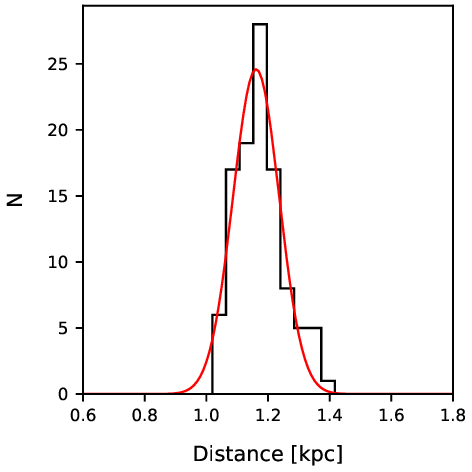}      
    \caption{Distance distribution of members. The bin size is about 0.04 kpc. The red curve represents the best-fit Gaussian distribution, with a prominent peak 1.16 kpc.}
    \label{fig7}
\end{figure}

The distance to NGC 2302 was derived from the 
inverse parallaxes of its members. Only members 
with parallaxes greater than five times the 
associated errors were used in this analysis. 
Figure~\ref{fig7} exhibits the distance 
distribution of the members. This distribution 
was fit to the Gaussian distribution. The distance 
to this cluster from the best-fit Gaussian is 
determined to be $1.16 \pm 0.08$ ($1\sigma$ dispersion) 
kpc. This results are consistent with the previous studies 
\citep{kharchenko2009,costa2015,bossini2019,
cantatanders2020,almeida2023,hunt2023}.

In general, the ages of clusters are estimated by 
comparing the observed CMDs with isochrones derived from 
theoretical stellar evolutionary models. In this study, we 
adopted the isochrones based on the PARSEC v2.0 tracks, 
incorporating stellar rotation ($\Omega/\Omega_{\text{crit}} = 0.3$) 
\citep{NCB25}. A grid of isochrones for solar metallicity 
was utilized. The total extinction, $A_V$, was calculated 
by multiplying the mean reddening ($\langle E(B-V)\rangle = 0.24$ 
by an $R_V$ value of 2.8. Extinction values for the $G_{BP}$ 
and $G_{RP}$ bands were determined using the relations 
described in the Appendix. Finally, the magnitudes and 
colors of the isochrones were adjusted for both the total 
extinction and a distance of 1.16 kpc. 

\begin{figure}
    \centering
    \includegraphics[width=80mm]{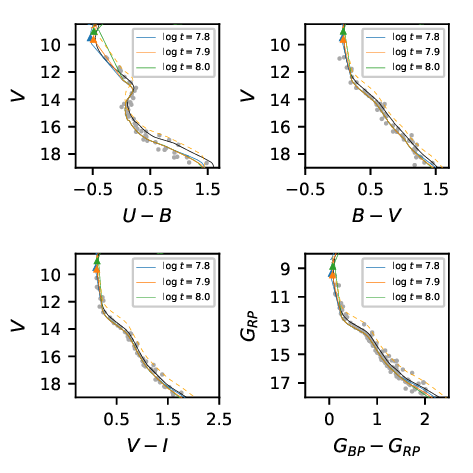} 
    \caption{CMDs of NGC 2302 derived from multi-color 
    photometry. Isochrones for $\log t = 7.8$ (blue), 7.9 (orange), and 8.0 (green) are shown for $Z = 0.008$. Additionally, the equal-mass binary sequence for $\log t = 7.9$ (orange dashed line) and a comparison isochrone with $Z = 0.015$ (black) for the same age are plotted to illustrate the effects of binarity and metallicity. Main-sequence turn-offs are indicated by triangles. The masses of stars near the main-sequence turnoff are approximately $5~M_{\odot}$.} 
    \label{fig8}
\end{figure}

Figure~\ref{fig8} displays the CMDs of the members along 
with isochrones for ages ranging from $\log t = 7.8$ to $8.0$. 
A comparison with the solar-metallicity isochrone (gray curve) 
at $\log t = 7.9$ reveals that, except in the $(V, V-I)$ CMD, 
the observed colors of late-type members are systematically bluer 
than predicted by the model at a given magnitude. This discrepancy 
is likely attributed to a reduced blanketing effect in stars 
with lower metallicity, as metallic absorption lines are heavily 
concentrated in the ultraviolet and blue bands. Consequently, we 
tested various metallicities to better match the colors of these 
late-type members and found that an isochrone with $Z = 0.008$ 
provides an excellent fit. 
As shown in Figure~\ref{fig8}, the adopted $Z = 0.008$ isochrone 
successfully reproduces the observed stellar colors across all CMDs.

The luminosity of the main-sequence turn-off MSTO is sensitive 
to the age of a cluster. However, this cluster lacks a well-defined 
MSTO, as no member stars have reached this evolutionary stage or 
evolved into red giants. In the $(V, V-I)$, $(V, B-V)$, and $(G_{RP}, 
G_{BP}-G_{RP})$ CMDs, the isochrones for different ages are closely 
spaced near the upper main sequence, making age discrimination 
challenging. In contrast, the $(V, U-B)$ CMD (Figure~\ref{fig8}) 
exhibits a wider dispersion in color, offering better sensitivity 
to age. Although the brightest observed members are systematically 
fainter than the predicted luminosities of main-sequence turn-off, 
their magnitudes and colors are generally consistent with the interval 
between the $\log t = 7.8$ and $8.0$ isochrones. Consequently, we 
constrain the age of this cluster to be between 60 Myr and 100 Myr 
($\log t = 7.9 \pm 0.1$, or $80 \pm 20$ Myr).

We derived the masses of individual members by adopting 
an isochrone with $\log t = 7.9$ and $Z = 0.008$. To account for 
the binary population observed in the CMDs, a grid of isochrones 
for the same age and metallicity was generated with mass ratios 
$q = m_s / m_p = 0.0$, $0.3$, $0.5$, $0.7$, and $1.0$, where 
$m_p$ and $m_s$ represent the primary and secondary masses, 
respectively. Individual primary masses ($m_p$) and mass ratios 
($q$) were determined by interpolating their magnitudes and colors 
within this binary isochrone grid. For systems with estimated $q < 0.3$, 
where the contribution of the secondary to the total light 
is negligible, we assigned the primary mass ($m_p$) as the stellar mass. 
For members fainter than the single-star ($q = 0.0$) isochrone, primary 
masses were obtained by interpolating their magnitudes using the 
mass-luminosity relation of main-sequence stars. Conversely, for members brighter than the 
equal-mass binary sequence ($q = 1.0$), their colors 
were interpolated accordingly to estimate their primary masses.

Summing these individual masses yields the total stellar 
mass of the cluster; however, this value represents 
a lower limit due to photometric incompleteness.
To address this, we derived the present-day mass 
function and fit it to the Kroupa IMF \citep{kroupa2001}.
Finally, the total stellar mass was estimated to 
be $333 \pm 48 \ M_{\odot}$ by integrating the mass 
function, with Poisson noise in each mass bin used 
to calculate the uncertainty. This cluster contains 
a total of $850 \pm 79$ members, extending down 
to the hydrogen-burning limit.

\begin{figure}[tb]
    \centering
    \includegraphics[width=80mm]{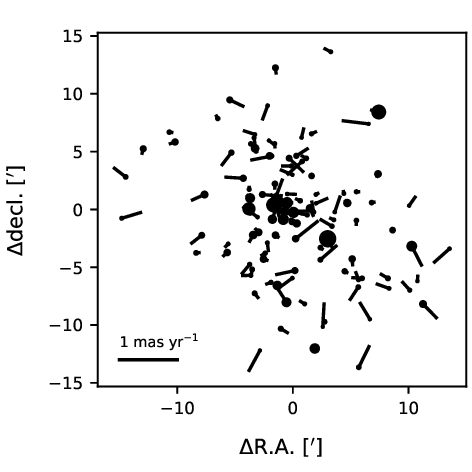}   
    \includegraphics[width=80mm]{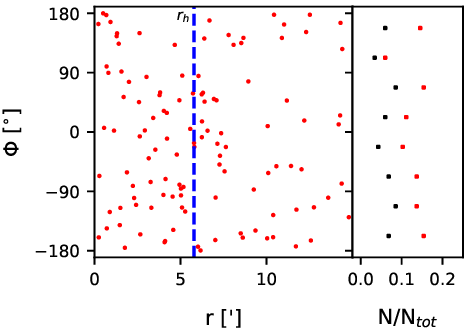}
    \caption{PM vectors (upper) and the $\Phi$ 
    distribution of members (lower). In the upper panel, 
    the black dots show the spatial distribution of members, 
    and the size of the dots is proportional to the brightness 
    of individual members. The solid lines represent the PM vectors 
    of individual members. The blue vertical line indicates 
    $r_h$ in the lower-left panel. In the lower-right panel, 
    the red and black dots show the histograms of $\Phi$ values 
    for all members and the members within $r_h$, respectively.} 
    \label{fig9}
\end{figure}

\section{Kinematics} \label{sec5}
We analyzed the PMs of individual members to 
investigate the kinematic properties of NGC 2302. 
The radial motion of stars in a stellar system 
induces two-dimensional apparent motions in the 
celestial plane. This projection effect varies with 
the distance to the system and its angular extent 
\citep{vL09}. We corrected for this effect 
on the PMs of cluster members using Equation 13 
of \citet{vL09}. For this correction, a median radial 
velocity of 27.6 km s$^{-1}$, derived from 15 members 
in the Gaia data, was adopted. Following the correction, 
the median PMs in R.A. and decl. are $-1.505$ mas yr$^{-1}$ 
and $-0.232$ mas yr$^{-1}$, respectively. 

Figure~\ref{fig9} displays the PM vectors of 
individual members. The orientations of the PM 
vectors are randomly distributed; in other words, 
the members do not exhibit any systematic 
motions, such as expansion, contraction, or 
streaming motion. To quantitatively analyze 
the PM vectors, we calculated the vectorial 
angle ($\Phi$) for individual members, defined 
as the angle between the position vector 
from the cluster center and the PM vector. 
A $\Phi$ of $0^{\circ}$ indicates that stars 
are moving away from the cluster center, while 
a $\Phi$ of $180^{\circ}$ indicates that stars 
are moving toward the center. 

\begin{figure}[tb]
    \centering
    \includegraphics[width=80mm]{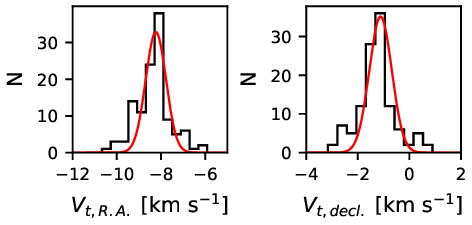}      
    \caption{Distributions of tangential velocity along R.A. and decl., respectively. The best-fit Gaussian distributions are represented as red curves. } 
    \label{fig10}
\end{figure}

The $\Phi$ distribution of individual members is 
presented in the lower-panel of Figure~\ref{fig9}. 
The $\Phi$ values are evenly distributed across 
all distances from the cluster center. The lower-right 
panel shows the fractional numbers relative to 
the total population at given $\Phi$ values. 
No significant pattern of expansion or 
contraction was found for either the entire 
member sample (red dots) or the subset within 
$r_h$ (black dots). These results indicate that 
the stellar motions within the cluster are isotropic.

\section{Dynamical State}\label{sec6}
This cluster exhibits a strong central concentration (see Section~\ref{sec4.1}) and lacks both tidal tails and streaming motions. Consequently, assuming a circular orbit, the Jacobi radius ($r_J$) of this cluster is expected to be similar to its tidal radius ($r_t$) for such an isolated system. We calculated $r_J$ using the following equation:

\begin{equation}
r_J = R_{\text{GC}}\left({M_{\text{cl}}\over 3M_{\text{MW}}}\right)^{1/3}
\end{equation}
\noindent where $R_{\text{GC}}$, $M_{\text{cl}}$, and $M_{\text{MW}}$ represent the galactocentric distance, the cluster mass, and the enclosed mass of the Galaxy, respectively. The Galactocentric distance of NGC 2302 ($R_{\text{GC}} = 8.9\text{ kpc}$) was calculated assuming a distance from the Sun to the Galactic center of $R_0 = 8.0\text{ kpc}$ (e.g., \citealt{2008ApJ...689.1044G}), based on the distance to this cluster ($d = 1.16\text{ kpc}$) and its Galactic coordinates $(l, b) = (221.96^\circ, -3.93^\circ)$. Adopting a rotational velocity of 220 km s$^{-1}$ in the solar neighborhood, the enclosed mass of the Galaxy within $R_{\text{GC}}$ is estimated to be $\sim 10^{11} \ M_{\odot}$. Given an $M_{\text{cl}}$ of $333 \ M_{\odot}$, $r_J$ is approximately 9.2 pc. By setting $r_t \approx r_J$, the concentration parameter ($c = \log r_t/r_c$) is found to be about 1.4. This value indicates a significantly stronger central concentration compared to typical open clusters ($c = 0.5 - 1.0$), suggesting that NGC 2302 is dynamically stable and likely mass-segregated.

We further examined the virial state of the cluster. 
The tangential velocities of members were computed 
by multiplying their PMs by the distance to the cluster. 
Since the spatial extent of the cluster is much smaller 
than its distance, individual distance variations among 
members do not introduce significant errors in their 
tangential velocities. 

Figure~\ref{fig10} displays the distribution of tangential 
velocities in R.A. and decl. The observed velocity dispersions ($\sigma_{\text{obs,R.A.}}$ and $\sigma_{\text{obs,decl.}}$) were derived 
from the standard deviation ($\sigma$) of the best-fit Gaussian distribution. These observed dispersions are a convolution of the intrinsic velocity dispersion ($\sigma_{\text{int}}$) and the observational error ($\sigma_{\text{err}}$), defined as $\sigma_{\text{obs}} = \sqrt{\sigma_{\text{int}}^2 + \sigma_{\text{err}}^2}$. After correcting for a mean tangential velocity error of $0.37 \ \mathrm{km \ s}^{-1}$, we derived intrinsic velocity dispersions of $\sigma_{\text{int,RA}} = 0.28 \ \mathrm{km \ s}^{-1}$ and $\sigma_{\text{int,decl}} = 0.23 \ \mathrm{km \ s}^{-1}$. The root-mean-square of these two values ($0.26 \ \mathrm{km \ s}^{-1}$) was adopted as the one-dimensional velocity dispersion of the cluster.

The virial velocity dispersion ($\sigma_{\text{vir}}$) was 
calculated using the equation from \citet{goodwin2006}:

\begin{equation}
\sigma_{\text{vir}} = \sqrt{{GM \over \eta R}}
\end{equation}

\noindent where $G$, $M$, $\eta$, and $R$ represent the gravitational constant, cluster mass, structure parameter, and cluster radius, respectively. We adopted a cluster mass of $333 \ M_{\odot}$. A concentration parameter of $c \approx 1.42$ corresponds to $\eta \sim 10$ \citep{portegies2010}. Adopting the half-number radius ($r_h = 1.9 \ \mathrm{pc}$) as the cluster radius yields an expected virial velocity dispersion of $\sigma_{\mathrm{vir}} \approx 0.27 \pm 0.02 \ \mathrm{km \ s}^{-1}$, which is in good agreement with the observed value ($0.26 \ \mathrm{km \ s}^{-1}$). To address the distinction between $r_h$ and the half-mass radius ($r_m$), we also directly calculated $r_m$ using individual member masses and mass ratios ($q \ge 0.3$), obtaining $r_m \approx 2.51'$ ($\approx 0.85 \ \mathrm{pc}$). Due to faint-end observational incompleteness, $0.85 \ \mathrm{pc}$ serves as a lower limit for $r_m$, whereas $r_h = 1.9 \ \mathrm{pc}$ acts as a conservative upper bound. Applying $r_m = 0.85 \ \mathrm{pc}$ gives $\sigma_{\mathrm{vir}} \approx 0.41 \ \mathrm{km \ s}^{-1}$. Considering that the true $r_m$ likely lies between these bounds, the derived range of $\sigma_{\mathrm{vir}}$ ($0.27 - 0.41 \ \mathrm{km \ s}^{-1}$) is consistent with the observed velocity dispersion. Furthermore, no significant signatures of kinematic expansion or contraction are detected in the internal velocity field. Combined with these kinematic signatures, we conclude that NGC 2302 is in virial equilibrium.

We also estimated the relaxation time ($t_{\text{relax}}$) of 
the cluster following \citet{binney2008}:
\begin{equation}
t_{\text{relax}} = \frac{0.1N}{\ln{N}}t_{\text{cross}},
\end{equation}
\noindent where $N$ and $t_{\text{cross}}$ are the number of 
members and the crossing time, respectively. Using $r_h$ 
and the observed velocity dispersion, $t_{\text{cross}}$ was 
calculated to be approximately 7.3 Myr. With $N = 850 \pm 79$ 
(Section 4.3), the relaxation time is estimated to be approximately 
$92 \pm 7 \ \mathrm{Myr}$ by propagating the uncertainty in $N$ 
into Equation (2), which is comparable to the age of the cluster 
($80 \pm 20 \ \mathrm{Myr}$).

When adopting the half-mass radius of $r_m \approx 0.85 \ \mathrm{pc}$, 
$t_{\text{cross}}$ decreases to approximately $3.3 \ \mathrm{Myr}$, 
yielding a shorter relaxation time of $t_{\mathrm{relax}} = 41 \pm 3 \mathrm{Myr}$. 
Depending on the adopted radius ($r_m$ or $r_h$), the derived 
relaxation time spans $\sim 41 - 92 \ \mathrm{Myr}$. Because this 
timescale is shorter than or comparable to the cluster age, and given 
that $r_m$ ($0.85 \ \mathrm{pc}$) is noticeably smaller than $r_h$ 
($1.9 \ \mathrm{pc}$), we conclude that NGC 2302 has reached a state 
of dynamical relaxation.

\begin{figure}[tb]
    \centering
    \includegraphics[width=80mm]{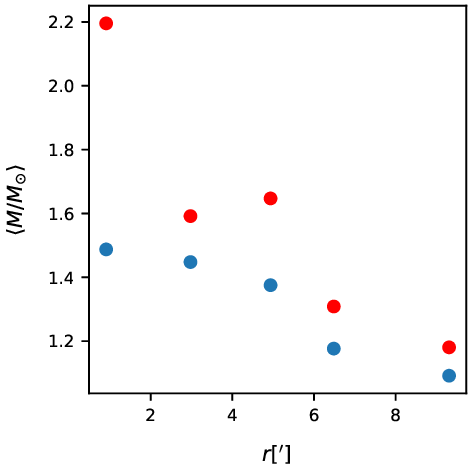}      
    \caption{Radial distribution of stellar masses. Red and blue dots are 
    mean and median stellar masses, respectively.}
    \label{fig11}
\end{figure}

These results collectively suggest that the cluster 
should exhibit mass segregation. To verify this, we 
investigated the radial distribution of stellar masses, 
as shown in Figure~\ref{fig11}. The mean stellar masses 
(red dots) are highest in the innermost region and decrease 
with increasing projected distance from the center. 
To minimize statistical fluctuations, we also examined 
the distribution of median stellar masses (blue dots). 
This radial mass distribution shows a smoother decline 
with distance, confirming that high-mass members are 
preferentially concentrated in the inner regions of 
the cluster.

\begin{figure}
    \centering
    \includegraphics[width=80mm]{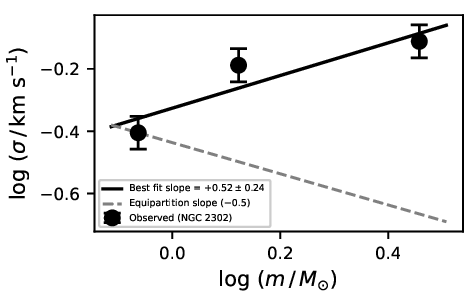}      
    \caption{Intrinsic 1D velocity dispersion ($\sigma$) versus mean stellar mass 
     for 104 member stars ($G_{RP} < 17$ mag) divided into three mass bins. The solid 
     line represents the linear fit ($+0.52 \pm 0.24$), and the dashed line 
     indicates the theoretical equipartition slope of $-0.5$.}
    \label{fig12}
\end{figure}

If the observed mass segregation in this cluster is indeed of 
dynamical origin, it should be accompanied by the process of 
kinetic energy equipartition ($\sigma \propto m^{-0.5}$). However, 
when testing for energy equipartition by dividing the 104 member stars 
($G_{RP} < 17$ mag) into three mass subsamples containing an equal 
number of stars ($\sim 35$ stars per bin), the observed slope between 
stellar mass and intrinsic 1D velocity dispersion departs significantly 
from the theoretical value of $-0.5$. As shown in Figure~\ref{fig12}, a linear 
fit yields a positive slope of $+0.52 \pm 0.24$. Instead of the expected 
decrease in velocity dispersion with increasing mass, the more massive 
stellar subsample exhibits a noticeably larger velocity dispersion 
than the lower-mass subsample, indicating that the cluster has not 
yet achieved full kinematic equipartition. While unresolved high-mass 
binary systems may contribute to this trend, this distinct kinematic 
feature suggests that the observed mass segregation is not purely 
dynamical, pointing toward the possibility of the in-situ formation 
of high-mass stars within the central region \citep{BD98,BB06}.

It is worth discussing how uncertainties in stellar mass estimates and unresolved binaries could potentially affect these conclusions. Systematic uncertainties arising from the adopted isochrone—such as slight variations in age, distance, or metallicity—would uniformly shift the mass scale of member stars, preserving the relative mass ordering and the overall kinematic trend. Regarding unresolved binaries, individual masses in this work were derived by interpolating observed photometry onto a grid of binary isochrones constructed for various mass ratios ($q$). Since the analyzed member stars are almost exclusively main-sequence stars, complex binary evolution pathways do not complicate our analysis, allowing our binary flux-addition grid to yield reliable mass estimates without systematic overestimation. Crucially, because our kinematic analysis relies solely on Gaia PMs without using radial velocities, orbital motions of unresolved binaries at a distance of 1.16 kpc are far too small to affect the proper motion dispersion. Consequently, our interpretation that NGC 2302 departs from kinetic energy equipartition ($\sigma \propto m^{+0.52}$) and supports an in-situ formation scenario remains physically robust against these uncertainties.

When strong dynamical interactions occur in the central regions, low-mass stars are expected to be ejected to the outskirts, which typically gives rise to radial anisotropy in the outer regions \citep{2020ApJ...899..121L}. However, as shown in Figure~\ref{fig9}, the distribution of PM vectorial angles shows isotropic stellar motions with no strong anisotropic signatures. The Jacobi radius of this cluster is approximately 9.2 pc. However, because the surface density of this cluster drops to the background level at an apparent radius of $15^{\prime}$ (equivalent to 5.1 pc), severe foreground and background star contamination makes it impossible to reliably track member stars beyond this point. This observational limitation prevents us from probing the cluster out to its Jacobi radius, introducing additional uncertainties regarding the overall physical size, the total mass of the cluster, and the kinematic behavior at the outer regions. Consequently, this unidentified outer halo limits a comprehensive understanding of the full dynamical status of this cluster.

\section{Summary} \label{sec6}
NGC 2302 is a valuable target for examining 
stellar dynamics. This cluster is estimated to 
contain approximately 850 members ($333 M_{\odot}$), 
of which 117 were identified as high-probability members 
using Gaia data. Our analysis, based on multi-color 
photometry and Gaia astrometry, yields the following results:

The mean reddening of the cluster is $\langle E(B-V)\rangle = 0.24 \pm 0.05$ (s.d.). The small standard deviation indicates negligible differential reddening across the field. Multi-color photometry from optical to near-infrared wavelengths yields a total-to-selective extinction ratio, $R_V$, of $2.8 \pm 0.1$, which is slightly lower than the Galactic average (3.0--3.1). By inverting member parallaxes, we determined the distance to NGC 2302 to be 1.16 kpc. Based on isochrone fitting on CMDs, the cluster age is estimated at $80 \pm 20$ Myr. Additionally, the relatively blue colors of late-type members compared to solar-neighborhood counterparts are consistent with a subsolar metallicity of $Z = 0.008$.

The cluster members are distributed within a cluster radius ($r_{\text{cl}}$) of 5.1 pc, with half the members located within a half-mass radius ($r_h$) of 1.9 pc. The radial surface density profile, fit to the \citet{king1962} model, yields a core radius ($r_c$) of $0.35 \pm 0.05$ pc. Its tidal radius ($r_J$ or $r_t$) is 
about 9.2 pc. These structural parameters indicate that 
the members are highly concentrated in the innermost region.

NGC 2302 appears kinematically stable; the PMs of its members 
show no significant patterns of expansion, contraction, 
or streaming motion. The observed velocity dispersion 
confirms that the cluster is in a virial state. The relaxation 
time is approximately 90 Myr, which is comparable to its age, suggesting that the cluster is currently approaching a state of dynamical relaxation. Furthermore, a clear signature of mass segregation was detected in the radial distribution of stellar masses.

However, because no definitive evidence of kinetic energy equipartition was found—with the higher mass stars instead exhibiting a larger velocity dispersion—the conventional pathway of dynamical relaxation cannot fully account for the internal kinematics of the cluster. This lack of equipartition suggests that the in-situ formation of high-mass stars within the central region remains a plausible scenario, likely preserving the unrelaxed primordial velocity structures of 
this cluster.


\acknowledgments
The authors thank the anonymous referee for constructive comments and suggestions. 
This work has made use of data from the European Space Agency (ESA) mission Gaia (https://www.cosmos.esa.int/gaia), processed by the Gaia Data Processing and Analysis Consortium (DPAC, https://www.cosmos.esa.int/web/gaia/dpac/consortium). Funding for the DPAC has been provided by national institutions, in particular the institutions participating in the Gaia Multilateral Agreement. 
The Digitized Sky Surveys were produced at the
Space Telescope Science Institute under U.S. Government
grant NAG W-2166. The images of these surveys are based on
photographic data obtained using the Oschin Schmidt Telescope 
on Palomar Mountain and the UK Schmidt Telescope.
The plates were processed into the present compressed digital
form with the permission of these institutions. 
The authors also acknowledge the use of the 61-inch Kuiper Telescope of the Steward Observatory, University of Arizona. 
This work was supported by the National Research Foundation 
of Korea (NRF) grant funded by the Korean government (MSIT; 
grant Nos. RS-2022-NR072247 and 2022R1C1C2004102) and the 
research grant of Kongju National University in 2026. This paper 
is dedicated to the memory of Yuna Lee, who led this research with 
great passion and is now a star in the night sky.


\appendix

\begin{figure}
    \centering
    \includegraphics[width=90mm]{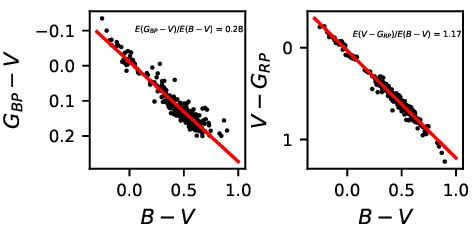}      
    \caption{Color-color relations of 248 early-type stars across 
    seven young open clusters. Red solid lines indicate the best-fit 
    linear relations, with slopes representing the reddening vectors 
    as labeled in the upper-right corner of each panel.} 
    \label{fig13}
\end{figure}

\section{Gaia Extinction Law}
We derived the relationship between the total-to-selective 
extinction ratio, $R_V$, and the color excesses 
in the Gaia passbands. To this end, we compiled $UBV$ 
photometric data for 248 early-type stars across seven 
young open clusters, covering a reddening range 
of $E(B-V) = 0.029$–1.221. These data were sourced from 
our previous studies: NGC 2353 \citep{sos1}, IC 1848 \citep{sos2}, 
IC 2162 \citep{sos5}, IC 1805 \citep{SBC17}, NGC 1624 \citep{sos4}, 
NGC 1893 \citep{sos3}, and NGC 2264 \citep{SBC08}. The $UBV$ 
photometry in these studies is well-calibrated to the Johnson-Cousins 
photometric system \citep{johnson1953}. The reddening law toward 
these clusters is consistent with the Galactic mean value ($R_V = 3.1$). 
Additionally, the $G_{BP}$ and $G_{RP}$ magnitudes for 
these stars were obtained from Gaia DR3 \citep{gaia2023}.

Figure~\ref{fig13} shows the observed 
color-color relations. The $G_{BP} - V$ and 
$V - G_{RP}$ colors increase linearly with 
$B-V$ for $E(B-V) \lesssim 1.0$. The slopes of these 
relations represent the reddening vectors, which were 
determined using the least-squares method: 
\begin{align}
    E(G_{BP} - V)/E(B-V) &=  0.28, \\
    E(V - G_{RP})/E(B-V) &= 1.17. 
\end{align}

Since the $R_V$ for the observed clusters 
is known to be 3.1, the relationships between $R_V$ and 
the color excess ratios are derived as follows:
\begin{align}
    R_V &= 11.1 E(G_{BP}-V)/E(B-V)\\&= 2.65 E(V-G_{RP})/E(B-V). 
\end{align}
These relations yield total extinction ratios of $A_{G_{BP}}/A_V = 1.09$ 
and $A_{G_{RP}}/A_V = 0.62$. These results are in good agreement with 
previous studies, such as $A_{G_{BP}}/A_V = 1.07$ and $A_{G_{RP}}/A_V = 0.62$ 
from \citet{CCM89}, and $A_{G_{BP}}/A_V = 1.08$ and $A_{G_{RP}}/A_V = 0.63$ 
from \citet{OD94}.

\bibliography{sosviii}






\end{document}